\documentclass{article}

\usepackage{graphicx} 
\usepackage{amsmath,amssymb}
\usepackage{multirow}
\usepackage{geometry}
\usepackage{url}

\usepackage[utf8]{inputenc}
\usepackage{authblk}
\usepackage{cite}

\title{The plan for a super $\eta$ factory at Huizhou accelerator complex}

\author[1,2]{Xurong Chen}
\author[1,2]{Xionghong He}
\author[1,2]{Qiang Hu}
\author[1,2]{Dexu Lin}
\author[1,2]{Yang Liu}
\author[1,2]{Hao Qiu}
\author[1,2]{Xu Sun}
\author[1,2]{Ye Tian}
\author[1,2]{Rong Wang}
\author[1,2]{Honglin Zhang}
\author[1,2]{Yapeng Zhang}
\author[1,2]{Chengxin Zhao}
\affil[1]{Institute of Modern Physics, Chinese Academy of Sciences, Lanzhou 730000, China}
\affil[2]{School of Nuclear Science and Technology, University of Chinese Academy of Sciences, Beijing 100049, China}

\date{May 2024}

\begin{document}

\maketitle

\begin{abstract}
As an approximate Goldstone boson with zero quantum number and zero standard model charge,
the decay processes of long-lived $\eta$ meson offer a unique
opportunity to explore new physics beyond the standard model and new sources
of CP violation, as well as test the low-energy QCD theory and measure the
fundamental parameters of light quarks. To pursue these goals in the physics
frontiers, we propose a plan to construct a super $\eta$ factory at HIAF
high-energy terminal or at CiADS after its energy upgrade.
The high-intensity proton beam at HIAF enables the production
of a vast number of $\eta$ samples,
exceeding $10^{13}$ events per year in the first stage,
utilizing multiple layers of thin targets made of light nucleus.
This paper presents the physics goals, the first-version conceptual design of
the spectrometer, and some preliminary simulation results.
\end{abstract}

\section{Introduction}
\label{sec:intro}

The high-luminosity frontier is one way to the new physics \cite{Proceedings:2012ulb},
as any small deviations from the standard model (SM) predictions
in the high-precision measurements are implications
of the new physics beyond the SM.
In the next decade, the upcoming high-intensity proton accelerator will
offer a unique opportunity for searching the new physics
at an unprecedented level. Actually, there are more and more signs of new physics,
such as the anomalous muon magnetic moment $(g-2)_{\mu}$
\cite{Muong-2:2023cdq,Muong-2:2021ojo,Muong-2:2021vma},
the X17 boson from the decay of the excited state of $^8$Be
\cite{Krasznahorkay:2015iga,Feng:2016jff,Feng:2016ysn},
the lepton flavor universality violation in bottom quark decays
\cite{LHCb:2021trn,LHCb:2021lvy,LHCb:2019efc,Alonso:2015sja,Patnaik:2023ins},
the excesses of the cosmic positron and electron
\cite{DAMPE:2017fbg,PAMELA:2008gwm,Chang:2008aa,HESS:2008ibn},
the narrow $\gamma$ ray emission from the galactic bulge \cite{Jean:2003ci},
and the unexplored natures of dark matter
\cite{Arbey:2021gdg,Oks:2021hef,Bertone:2018krk,Young:2016ala,Feng:2010gw}
and dark energy \cite{Frieman:2008sn,Vagnozzi:2021quy,Joyce:2016vqv,Li:2012dt,Arun:2017uaw}.
Up to now, physicists do not observe any new physics in the
high-energy frontier with the large hadron collider.
Therefore, some people argue that the new physics of the
hidden sector is at low energies \cite{Batell:2009di,Lanfranchi:2020crw},
but it is faintly coupled to SM matter, which makes it elusive.
For example, the production rates of the light portal particles
connecting the hidden sector with the SM sector are magnitudes higher at low energies \cite{Batell:2009di}.
Moreover, for the low-energy fixed target experiment,
the luminosity can be much higher with a thick target.

The $\eta$ meson is of particular interests, as it is an approximate Goldstone boson
from spontaneous chiral symmetry breaking and with zero SM charge \cite{ParticleDataGroup:2022pth}.
Many strong and electromagnetic decay channels of $\eta$ are forbidden at the leading order,
which relatively enhance the rare decay channels of $\eta$ meson
that are sensitive to new physics.
The $\eta$ meson is a wonderful low-energy lab for searching the new physics beyond
the SM, via observing the dark portal particles from $\eta$ decays
\cite{REDTOP:2022slw,Batell:2009di,Lanfranchi:2020crw}
or measuring the small discrete symmetry breaking such as CP violation
and charged lepton flavor violation \cite{Gan:2020aco,Gardner:2019nid}.
A thorough discussion on the theoretical developments of $\eta$ and $\eta^{\prime}$
decays were impressively given in a recent review article \cite{Gan:2020aco},
regarding the high-precision tests of fundamental physics.
There are four kinds of portals predicted by the latest theoretical models:
the vector portal \cite{Holdom:1985ag,Galison:1983pa,Fayet:1990wx,Fayet:1980rr},
the scalar portal
\cite{Burgess:2000yq,OConnell:2006rsp,Batell:2018fqo,Batell:2017kty,Patt:2006fw,Silveira:1985rk,Pospelov:2007mp},
the axion-like portal \cite{Georgi:1986df,Bauer:2017ris,Aloni:2018vki,Landini:2019eck,Ertas:2020xcc}
and the heavy neutral lepton portal \cite{Gorbunov:2007ak,Atre:2009rg,Gninenko:2009ks}.
The portal particles are important in theories for connecting
the Dark Sector and the SM Sector.
All these light portal particles can be accessed from the rare decays
of $\eta$ meson \cite{Gan:2020aco,REDTOP:2022slw}.
The symmetry and symmetry breaking are at the heart of modern physics.
Finding new sources of CP violation is essential to explain the baryon-antibaryon
asymmetry in the world where we live in. Any charged lepton flavor violation is a strong
indication of beyond SM physics. Many $\eta$ decay channels provide the precise tests
of these symmetry breakings. The precise measurements of $\eta$ decay channels are critical
for us to understand the C, P, T, CP, and charged lepton flavor violations.

In addition to the new physics searches, the high-precision study of $\eta$
decay provides a unique way to test the quantum chromodynamics (QCD) theory at low energies
\cite{Weinberg:1978kz,Gasser:1983yg,Gasser:1984gg,Ecker:1994gg,Pich:1995bw,Bernard:2006gx},
to probe the $\eta$ structure
\cite{A2:2013wad,Adlarson:2016hpp,Pszczel:2019sdk,KLOE:2011qwm,CELSIUSWASA:2007ifz,BESIII:2015zpz,Escribano:2015nra},
to precisely measure the mass difference of light quarks
\cite{Gell-Mann:1968hlm,Weinberg:1977hb,Dashen:1969eg,Gasser:1984pr,Kaplan:1986ru,Leutwyler:1996qg},
to verify the axial anomaly \cite{Wess:1971yu,Witten:1983tw,Bernstein:2011bx}.
From the electromagnetic decay channels associated with virtual and real photons,
the $\eta$ transition form factor can be constrained with much smaller uncertainties
\cite{A2:2013wad,Adlarson:2016hpp,Pszczel:2019sdk,KLOE:2011qwm,CELSIUSWASA:2007ifz,BESIII:2015zpz},
which eventually helps us better understanding the muon anomalous magnetic moment
\cite{Muong-2:2023cdq,Muong-2:2021ojo,Muong-2:2021vma}.
The quark masses are the fundamental parameters of the SM.
Measuring the isospin-breaking 3$\pi$ decay channels of $\eta$ is one important way
to experimentally constrain the light quark masses. The high-precision measurements
at a super $\eta$ factory will reduce the uncertainties of QCD parameters significantly.
The precise measurements of some $\eta$ rare decays provide a strict test
of chiral perturbation theory at high orders \cite{Jetter:1995js},
which is a rigorous and effective theory for the strong interaction at low energies.

As the $\eta$ meson decays involve such plentiful physics,
there are already many measurements of $\eta$ decays with the current
facilities worldwide. First, there are the hadronic productions of $\eta$ from
fixed-target experiments, such as the WASA-at-COSY experiment
\cite{WASA-at-COSY:2018jdv,WASA-at-COSY:2014wpf,Husken:2019dou}
and the LHCb experiment \cite{LHCb:2023iyw,LHCb:2016hxl}.
The WASA-at-COSY collaboration used the proton beam at COSY
and an internal pellet target, and the statistic of $\eta$ events is at the level of $10^8$.
Second, the radiative decays of $\phi$ and $J/\psi$ at electron-positron colliders
produced a decent amount of $\eta$ samples of low background.
The number of $\eta$ events from $\phi$ decay by KLOE collaboration is at the level of $10^8$
\cite{Krzemien:2019ktq,KLOE-2:2020ydi,KLOE-2:2016zfv,KLOE:2008arm},
while the number from $J/\psi$ decay by BESIII collaboration is at the level of $10^7$
\cite{BESIII:2021fos,BESIII:2015fid,BESIII:2022tas,BESIII:2019gef,BESIII:2017kyd}.
Third, the photoproduction experiment is also a clean way to produce $\eta$ meson,
such as the A2 experiment at MAMI \cite{A2:2017gwp,A2:2013wad,CrystalBallatMAMI:2010slt},
and the JLab Eta Factory (JEF) \cite{Gan:2020aco,PR12-14-004}
by exploiting the Primakoff effect \cite{Primakoff:1951iae}.
JLab has a long tradition of studying the neutral meson physics
via the Primakoff reaction \cite{PrimEx:2010fvg,PrimEx-II:2020jwd}.
For JEF experiment, around $10^9$ tagged $\eta$ events will be collected in years,
with the GlueX spectrometer \cite{GlueX:2020idb,Asaturyan:2021ese}.
Thanks to the high-performance calorimeter
and the high energy of the incident photon up to 11 GeV,
the backgrounds are well in control for the neutral decay channels.
The JEF experiment has the advantages in measuring
the neutral decay channels of $\eta$ meson.

Aiming at the intriguing discovery potentials of light dark portal particles
and the strong tests of the standard model,
it is imperative to build a super $\eta$ factory with the high-intensity accelerators,
to acquire the unprecedented $\eta$ meson samples.
To pursue a vast number of $\eta$ events,
REDTOP (Rare Eta Decays To Observe Physics beyond the standard model)
experiment \cite{REDTOP:2022slw} is proposed
for the 2021 US community study on the future of particle physics,
with some novel detector techniques.
In China, a High-Intensity heavy-ion Accelerator Facility (HIAF) is under construction
at Huizhou city, by Institute of Modern Physics (IMP), Chinese Academy of Sciences (CAS),
which is competitive in the beam intensity. With this near-future infrastructure,
we suggest a super $\eta$ factory at HIAF high-energy terminal.
There is no doubt that the proposed Huizhou $\eta$ factory will bring many
fruitful physics results and it will push the accelerator and detector
technologies to a much higher level.

The organization of paper is as follows.
The suggested Huizhou $\eta$ factory and its physics goals are introduced in Sec. \ref{sec:physical-goals}.
The conceptual design of the spectrometer is present in Sec. \ref{sec:detector}.
Some preliminary results of the simulations are shown in Sec. \ref{sec:simulation},
for some golden channels of the experiment.
In Sec. \ref{sec:summary}, a concise summary and the outlooks are given.

\section{Huizhou $\eta$ factory and its goals}
\label{sec:physical-goals}

\begin{figure}[htbp]
\begin{center}
  \includegraphics[width=0.75\textwidth]{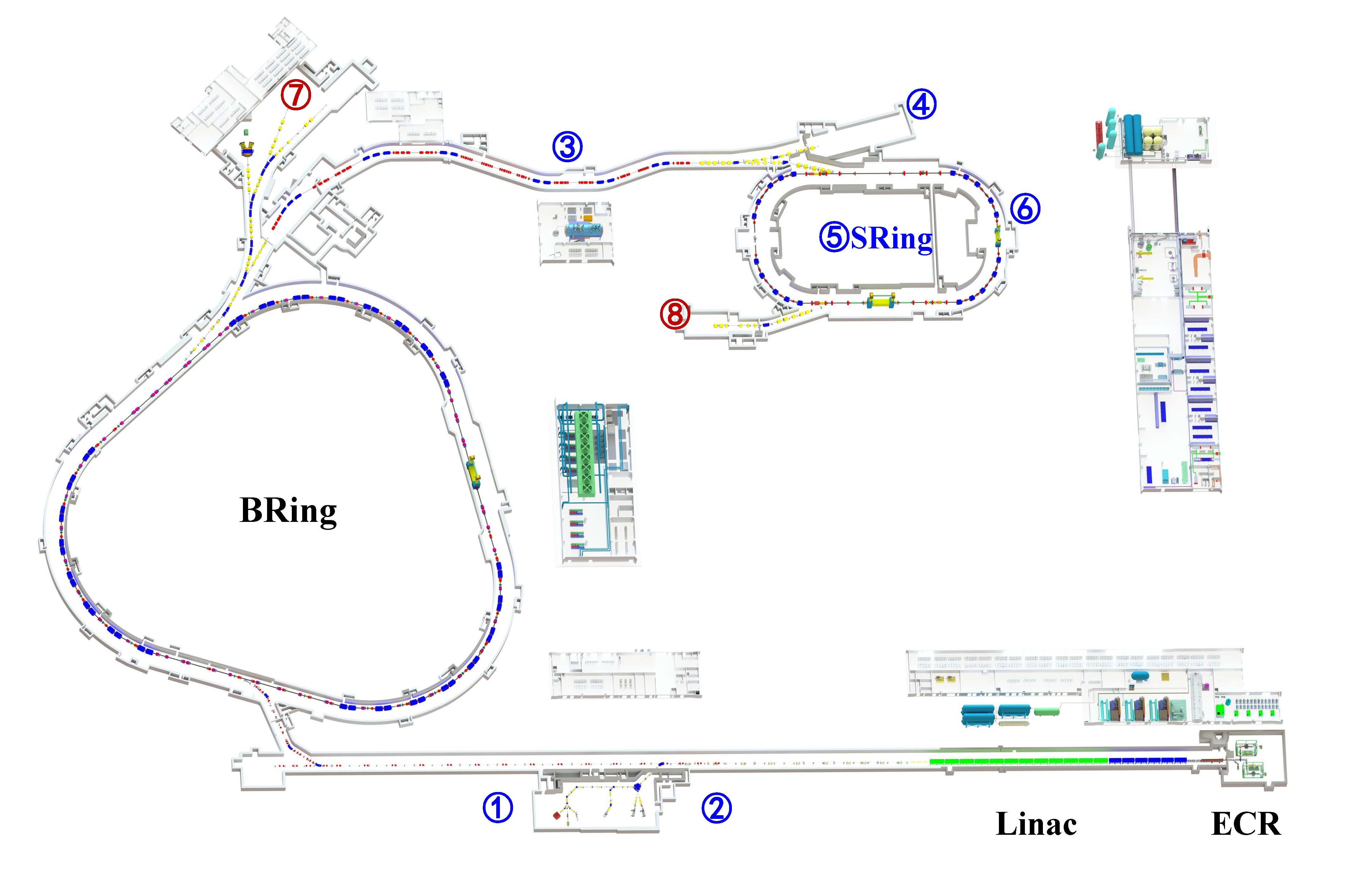}
  \caption{
    The layout of HIAF facility. The number ``\textcircled{\small 7}'' indicates
    where the high-energy multidisciplinary terminal locates.
  }
  \label{fig:HIAF-layout}
\end{center}
\end{figure}

\begin{table}[h]
\centering
  \caption{The list of the main physics goals of Huizhou eta factory.}
    \renewcommand\arraystretch{1.5}
	\begin{tabular}{|c|c|c|}
      \hline\hline
        \multicolumn{2}{|c|}{Physics goals} & Decay channel\\
      \hline
      \multirow{9}{*}{New physics} & Dark photon \& X17 & $e^+e^-\gamma$ \\
      \cline{2-3}
                & \multirow{2}{*}{Dark higgs}    & $\pi^+\pi^-\pi^0$ \\
                &                                & $\pi^0e^+e^-$ \\
      \cline{2-3}
            & \multirow{2}{*}{Axion-like particle}   & $\pi^+\pi^-e^+e^-$ \\
            &                                        & $\pi^+\pi^-\gamma\gamma$ \\
      \cline{2-3}
                & \multirow{2}{*}{CP violation}  & $\pi^+\pi^-\pi^0$ \\
                &                                & $\pi^+\pi^-e^+e^-$ \\
      \cline{2-3}
            & \multirow{2}{*}{Lepton flavor violation} & $\gamma\mu^+e^-$ / c.c. \\
            &                                          & $\mu^+e^-$ / c.c. \\
      \hline
      \multirow{10}{*}{Precision test of the SM} & \multirow{3}{*}{$\eta$ transition form factor} & $e^+e^-\gamma$ \\
                &                                & $e^+e^-e^+e^-$ \\
                &                                & $\pi^+\pi^-\gamma$ \\
      \cline{2-3}
                & \multirow{2}{*}{Light quark masses}  & $\pi^+\pi^-\pi^0$ \\
                &                                      & $\pi^0\pi^0\pi^0$ \\
      \cline{2-3}
                & \multirow{2}{*}{Chiral anomaly}& $\gamma\gamma$ \\
                &                                & $\pi^+\pi^-\gamma$  \\
      \cline{2-3}
                &    Beyond SM weak decay        & $e^+e^-$ \\
      \cline{2-3}
                & \multirow{2}{*}{Test chiral perturbation theroy} & $\pi^+\pi^-\gamma\gamma$  \\
                &                                                  & $\pi^0\gamma\gamma$ \\
      \hline\hline
	\end{tabular}
  \label{tab:physics-goals}
\end{table}

The HIAF is an under-construction
major national science infrastructure facility of China,
located in Huizhou city, Guangdong province
in Southern China \cite{Yang:2013yeb,Yang:2021cpq,Zhou:2022pxl}.
The construction of HIAF started from December of 2018,
and it is going to be ready for commissioning at the end of 2025.
The HIAF is an accelerator complex mainly consisting of a superconducting
electron-cyclotron-resonance ion source, a continuous-wave superconducting ion linac, a booster synchrotron,
a high-energy fragment separator, and a high-precision spectrometer ring.
The layout of HIAF is depicted in Fig. \ref{fig:HIAF-layout}.
Many terminals are designed along the accelerator complex for experiments and applications.
With the high-intensity technology, HIAF is not only a powerful infrastructure
for the frontier studies in nuclear physics, high energy density physics, and atomic physics,
but also an excellent platform for heavy-ion applications in life, material and space sciences \cite{Zhou:2022pxl}.
HIAF is featured with delivering the unprecedented intense ion beams
from hydrogen to uranium with energy up to GeV/u.
The maximum energy for the proton beam is 9.3 GeV \cite{Yang:2013yeb,Yang:2021cpq,Zhou:2022pxl}.
With the heavy-ion beam, HIAF provides an extraordinary platform for
the studies of hypernuclei and the phase structure of high-density nuclear matter;
With the high-energy proton beam, HIAF gives an excellent opportunity
for studying light hadron physics and building a $\eta$ factory.

At HIAF, the intensity of the proton beam is higher than $10^{13}$ ppp (particles per pulse),
and the kinematic energy of the proton can reach 9 GeV through
the accelerations of the ion linac and the booster ring \cite{Yang:2013yeb,Yang:2021cpq,Zhou:2022pxl}.
The pulse rate is around several Hz.
A super $\eta$ factory is suggested to be build at the high-energy multidisciplinary terminal
after the booster ring, the terminal ``\textcircled{\small 7}'' shown in Fig. \ref{fig:HIAF-layout}.
The target is made of multiple foils of light nucleus ($^7$Li or $^9$Be)
with 1 cm gaps, which significantly reduces the coincident background from the same vertex
with no decline of the luminosity at the same time.
With the proton beam and light nuclear target,
the $\eta$ meson is produced efficiently with controlled background at HIAF.
The beam-energy thresholds are 1.26 GeV and 2.41 GeV
for $\eta$ and $\eta^{\prime}$ productions respectively.
In the proton-proton scattering and at beam energy of 1.8 GeV,
the cross section of $\eta$ meson production is large
\cite{Wilkin:2016mfn,Moskal:2003gt,Petren:2010zz,Calen:1998vh},
judged by the previous COSY data (around 100 $\mu$b) \cite{Wilkin:2016mfn}.
For the nuclear target, the cross section for $\eta$ meson production is even higher.
Based on the HIAF beam intensity and a lithium target of 1 cm thickness in total,
the luminosity is above $10^{35}$ cm$^{-2}$s$^{-1}$ for the Huizhou eta factory experiment.
Regardless of the capabilities of the detector and data acquisition systems,
the $\eta$ production rate can be higher than $10^{8}$ s$^{-1}$
on a light nuclear target ($>10^{15}$ per year).

China initiative Accelerator Driven Sub-critical system (CiADS) is another
high-intensity proton accelerator designed for the verification
of the principle of nuclear waste disposal
\cite{he:ipac2023-fryg2,Wang:IPAC2019-MOPTS059,liu:hiat2018-weyaa01,Liu:2019xgd,Wang:2024joa,Cai:2023caf}.
It provides a much powerful continuous proton beam.
The designed full power of CiADS accelerator is 2.5 MW,
with the beam intensity of $3.15\times 10^{16}$ pps.
The CiADS is also a good choice for building a super $\eta$ factory,
so long as the energy of CiADS is upgraded to about 2 GeV.
Since the upgrade of CiADS accelerator is anticipated to
take many years, the proposed Huizhou $\eta$ factory will
be located mainly at HIAF high-energy terminal.

At Huizhou $\eta$ factory, there is going to be a huge statistic of $\eta$ meson samples,
about four magnitudes more than the current $\eta$ events achieved worldwide.
With such an enormous yield of $\eta$ mesons, the main physical goals of
Huizhou $\eta$ factory are to discover the new physics via the search for new
particles and the discrete symmetry breaking, and to study the SM
with a very high precision. The new particles of interests from $\eta$
and $\eta^{\prime}$ decays are the predicted light portal particles below GeV
which connect faintly the SM sector with the Hidden sector,
such as the dark vector particles \cite{Holdom:1985ag,Galison:1983pa,Fayet:1990wx,Fayet:1980rr},
the dark scalar particles
\cite{Burgess:2000yq,OConnell:2006rsp,Batell:2018fqo,Batell:2017kty,Patt:2006fw,Silveira:1985rk,Pospelov:2007mp},
and the axion-like particles \cite{Georgi:1986df,Bauer:2017ris,Aloni:2018vki,Landini:2019eck,Ertas:2020xcc}.
The protophobic X17 boson of the fifth force \cite{Krasznahorkay:2015iga,Feng:2016jff,Feng:2016ysn}
also can be studied via rare $\eta$ decay.
The huge amount of $\eta$ mesons provides us a good opportunity to study
new sources of CP violation, which is essential for the matter-antimatter asymmetry in the universe,
and to search for the charged lepton flavor violation,
which is a clear and strong sign of new physics.
Last but not least, the precise measurements of $\eta$ decays are critical
for high precision study of the standard model,
such as strictly constraining the light quark mass difference,
precise measurement of meson structure, and high-precision test of chiral perturbation theory.
The main physics interests are listed in Table \ref{tab:physics-goals}.
Since the spectrometer of Huizhou eta factory excels particularly in
measuring the charged particles, the charged decay channels are the golden
channels of high priority for the proposed experiment, such as
$\eta\rightarrow \pi^+\pi^-\pi^0$,
$\eta\rightarrow e^+e^-\gamma$,
$\eta\rightarrow \pi^+\pi^-e^+e^-$,
and $\eta\rightarrow e^+e^-$.

\section{A compact and large-acceptance spectrometer with silicon pixels}
\label{sec:detector}

\begin{figure}[htbp]
\begin{center}
  \includegraphics[width=0.47\textwidth]{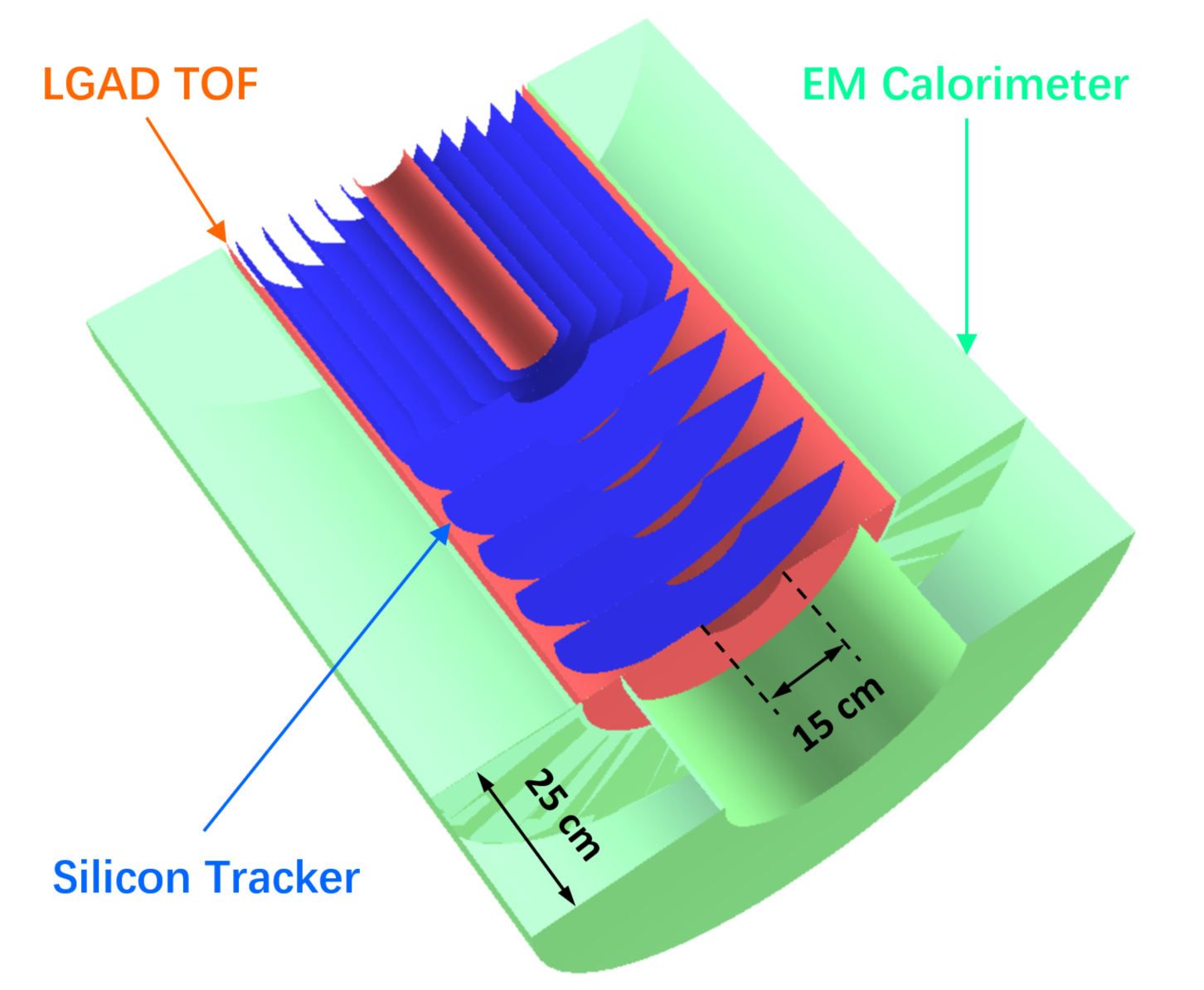}
  \caption{
    The conceptual design of a compact spectrometer for $\eta$ factory.
    The spectrometer mainly relies on the silicon detector technology,
    with the monolithic silicon pixel tracker and the fast LGAD TOF detector of low material budget.
    The silicon tracker is wrapped with a fast lead-glass calorimeter for high-energy photons.
  }
  \label{fig:Spectrometer}
\end{center}
\end{figure}

\begin{figure}[htbp]
\begin{center}
  \includegraphics[width=0.47\textwidth]{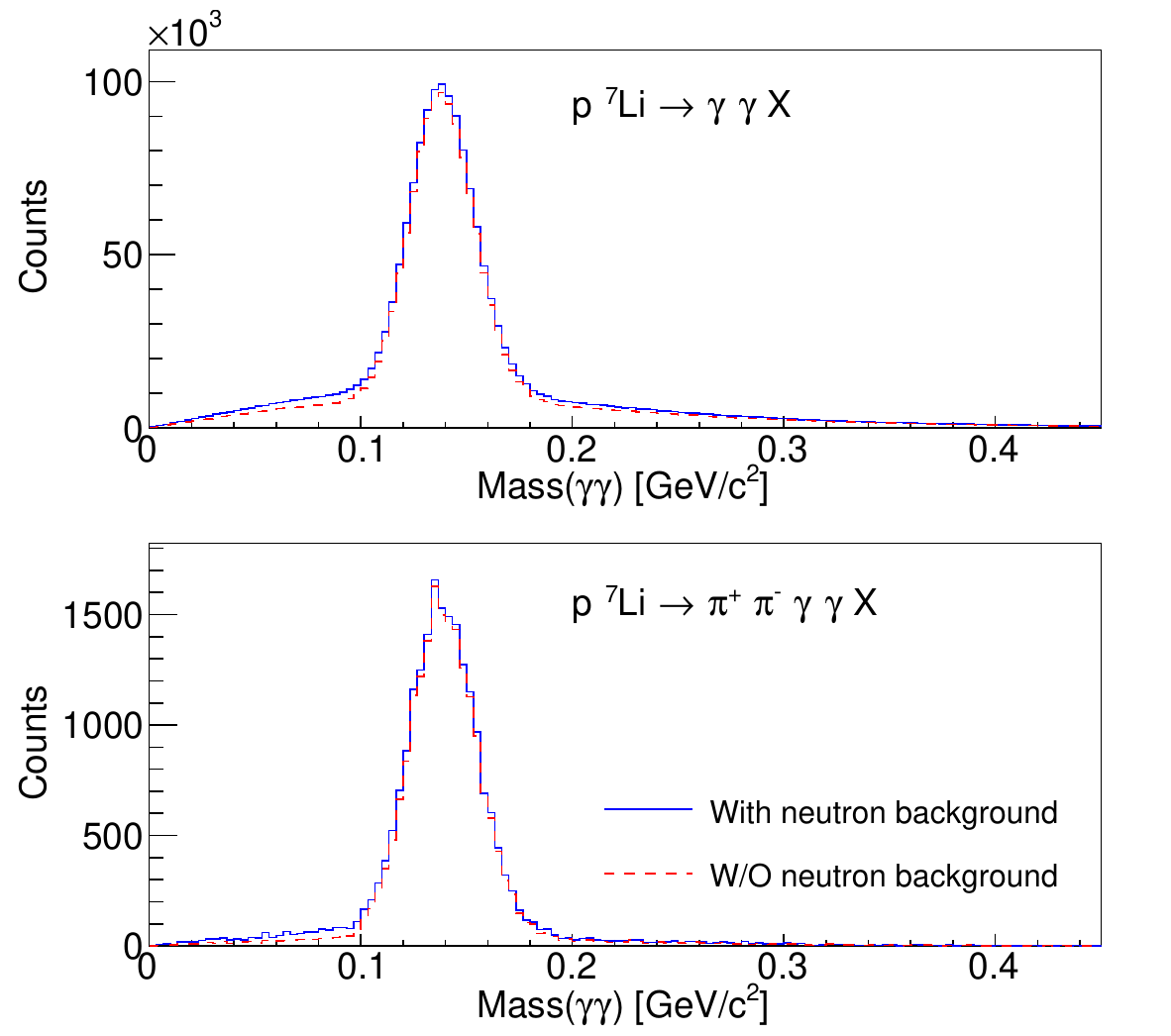}
  \caption{
    The invariant mass distributions of two $\gamma$'s from the simulations
    with and without the neutron contamination.
    The p-$^7$Li events are generated with GiBUU package.
    The $\gamma$'s are detected under two different situations:
    (1) we assume that the calorimeter can NOT distinguish the neutron from the photon (with neutron background);
    (2) we assume that the calorimeter can well distinguish the neutron from the photon (without neutron background).
  }
  \label{fig:background_of_pi0}
\end{center}
\end{figure}

With the fast development of the monolithic silicon pixel technology \cite{He:2023svm},
we came up with an idea of large acceptance and compact spectrometer with silicon pixels
to detect the final-state particles at a high event rate.
The current design of the spectrometer include four main parts:
the tracking system for charged particles made of silicon pixels,
the time of flight detector for particle identification
made of silicon Low-Gain Avalanche Detector (LGAD),
the Electro-Magnetic calorimeter (EM calorimeter) for photon measurement
made of lead glass \cite{APPEL1975495}, and a super-conducting solenoid.
The 3D design of the spectrometer is shown in Fig. \ref{fig:Spectrometer}.
Thanks to the high granularity and small position resolution of
the silicon pixel detector, it is a quite compact spectrometer
with a small volume. Therefore the EM calorimeter and the solenoid are
of small size, which reduce the cost for fabricating the spectrometer.
The inner radius of the super-conducting solenoid is just
about 70 centimeters and all the main detectors are inside the solenoid.

The multi-layer target is put inside the spectrometer
close to the entrance, so that to have a large acceptance
for the fixed-target experiment.
With the current conceptual design of spectrometer,
all forward particles except the small-angle particles are covered without dead zones.

To achieve a high-rate capacity of the silicon pixel tracker,
the silicon detector group has tried the dual measurements of
the energy and the arrival time of each pixel
\cite{Ren:2022ash,Yang:2021baf,Yang:2022qvv,Yang:2022iei,Huang:2023tcf}.
With the different arrival times, the hits from different events can be distinguished.
For the first-version of the silicon pixel chip ($\sim 1\times 1$ cm$^2$) of about 100k pixels,
it takes about 600 $\mu$s to read all the pixels in the scan mode,
and the resolution of arrival time is about 400 ns.
The objectives of the future silicon pixel chip are the resolution of 10 ns
for the arrival time, the pixel size of 30 $\mu$m,
and the scan time of 100 $\mu$s for about 100k pixels.
Under the particle multiplicity of Huizhou eta factory and with
the pixel chip more than 5 cm away from the interaction point,
the designed silicon pixel chip can easily record the events at the
event rate of more than 100 MHz.

For the current conceptual design of the spectrometer,
the material for the calorimeter is lead glass which generates only
the prompt Cherenkov photons. Therefore it has a good time resolution
around 100 ps for the particle detection.
At the same time, the lead glass is not sensitive
to the hadronic shower initiated by nucleons and pions, which means it has a small efficiency for neutron background and offers additional hadron background suppression capability.
Our Geant4 simulation \cite{GEANT4:2002zbu,Allison:2006ve,Allison:2016lfl}
finds that the low-energy neutrons
($E_{k}<0.3$ GeV) generate almost no hit in the lead glass calorimeter,
and the high-energy neutron (($E_{k}>1$ GeV) has just about 45\%
probability to deposit more than 10 MeV energy in the calorimeter.
As most of the neutrons from $pA\rightarrow X$ collision are low energy
neutrons, the neutron background in the photon measurement can be eliminated
effectively with the lead glass calorimeter. With the inelastic events
generated with GiBUU package \cite{Buss:2011mx,GiBUU-website,Gaitanos:2007mm,Weil:2012ji},
we show in Fig. \ref{fig:background_of_pi0}
the invariant mass distributions of two photons, with and without
the neutron background. The background for the $\pi^0$ signal
due to the neutron contamination is negligible,
especially for the channel $p{\rm ^7Li}\rightarrow \pi^+\pi^-\pi^0 X$.
In the simulation, the threshold for a hit in the calorimeter
corresponds to the signal generated by a 50 MeV photon.
The neutron deposits less energy in the calorimeter comparing to the photon, and with the same amount of energy deposition,
the hadronic shower initiated by the neutron generates less Cherenkov photons.
That is why the abundant neutron background at low energy is strongly suppressed
in the measurement of photon and $\pi^0$.

Although lead glass calorimeter is effective at suppressing hadron backgrounds and is cost-effective,
it has significant drawbacks compared to conventional crystal calorimeters.
Firstly, the low Cherenkov light yield and severe light attenuation
of lead glass result in poor energy resolution.
Lead fluoride crystals, which have less light attenuation, can be used instead,
but they are much more expensive. Another disadvantage is the poor radiation resistance.
While radiation-resistant types of lead glass can be used to improve this,
they suffer from worse light attenuation. Additionally,
due to the low light yield, the Cherenkov light being mainly in the UV range,
and the detector being in a magnetic field, large-sized UV-sensitive SiPMs are required.
These drawbacks present challenges for the usage of lead glass in this project.

One option is to use the ADRIANO2 \cite{adriano2} dual-readout calorimeter
currently being developed by the REDTOP group.
This design combines scintillation material and lead glass
to capture both Cherenkov light and scintillation light signals.
It employs longitudinal layering and readout, providing excellent energy resolution
and additional capability for low-energy particle identification.
This design addresses the shortcomings of using lead glass alone.

With the application of full silicon tracker of small pixel size,
the momenta of charged particles are measured very precisely with a high event rate,
and the sizes of all detectors scale down, which all depends on the size of inner tracker.
It is a compact spectrometer with a large acceptance for the fixed-target experiment,
and with competitive functions. The LGAD detector for time of flight measurement
has a small time resolution and a very low material budget.
The lead glass calorimeter is effective in reducing the neutron background,
but its energy resolution is not so good.
We also look for the new technologies of EM calorimeter which is capable
to work in a high event rate environment.
Therefore, with the current design of the spectrometer for Huizhou eta factory,
we focus more on the charged decay channels of $\eta$ meson.
The radiation dosage for the spectrometer is simulated with
both Geant4 \cite{GEANT4:2002zbu,Allison:2006ve,Allison:2016lfl}
and FLUKA \cite{Battistoni:2015epi,fluka.cern,fluka.org}.
Under the circumstance of 100 MHz of inelastic scattering rate
over a one-month data-taking period,
the innermost LGAD is expected to experience a maximum 1 MeV neutron equivalent fluence
of $3\times10^{12}~n_{\rm eq}/{\rm cm^{2}}$ and a maximum dose of 200 Gy.
Meanwhile, the lead glass of EM calorimeter is expected to experience
a maximum 1 MeV neutron equivalent fluence
of $5\times10^{11}~n_{\rm eq}/{\rm cm^{2}}$ and a maximum dose of 100 Gy.
Thus, these sub systems can survive several years
before significant radiation damage occurs.

\section{Some preliminary results of simulations}
\label{sec:simulation}

To see the physics impact and the feasibility of the experiment,
we have performed some simulation studies of some golden channels
for the Huizhou eta factory project. The simulation study is the
first step for us to acquire the details regarding the resolutions,
the efficiency of the signal channel, the background distribution,
the precision of the planed measurement,
and (or) the sensitivity to the new physics.

For the background events in p-A collisions,
we use GiBUU event generator \cite{Buss:2011mx,GiBUU-website,Gaitanos:2007mm,Weil:2012ji}
to do the simulation.
GiBUU is suitable for the proton-induced nuclear reactions from
low energy to intermediate energy, with the final-state interactions well handled \cite{Buss:2011mx}.
GiBUU event generator is based on the dynamical evolution of a colliding nucleus-nucleus system
within the relativistic Boltzmann-Uehling-Uhlenbeck framework,
which takes the hadronic potentials, the equation of state of nuclear matter,
and the collision terms into account.
In GiBUU, the low-energy collision is dominated by the resonance processes,
while the high-energy collision is described with a string
fragmentation model implemented in Pythia.
For the $\eta$ production, $N^{*}(1720)$ in the process
$NN\rightarrow NR$ plays a dominant role \cite{Lu:2015pva}.
Thus, GiBUU event generator covers perfectly the kinematical regions
of HIAF and CiADS accelerator facilities.

In our simulation, the kinematic energy of the proton beam is 1.8 GeV,
which is just below the $\rho$ meson production threshold
to lower the background.
With the lithium target, we find that the number of neutrons is about
one thousand times of the number of $\eta$ mesons,
and the number of $\pi^0$ mesons is about 50 times of that of $\eta$ mesons.
The decay chains of $\pi^0$ and $\eta$ are further coded by us.
For the signal event generations of the dark portal particles,
we composed a simple event generator for the channels of interests.
We also use another BUU generator \cite{Li:2001xh} and
the Urqmd package \cite{Bass:1998ca,Bleicher:1999xi,urqmd-website}
to estimate the $\eta$ production cross section.
The $\eta$ production probability is 0.76\% in the inelastic collisions.

To quantify the detection efficiency and the resolutions,
we have developed a detector simulation package ChnsRoot,
which is based on the FairRoot framework \cite{Al-Turany:2012zfk,fairroot.gsi.de}.
For now, we have the reliable fast simulation tool,
which is based on the parametrizations validated by Geant4 simulations.
The inner-most and outer-most radii of the silicon pixel tracker are
7.5 cm and 27.5 cm respectively.
The magnetic field strength is 0.8 Tesla.
The energy resolution of the calorimeter is $\delta(E)/E=\sqrt{a^2+b^2/(E/{\rm GeV})}$
for the photon, with $a=0.028$ and $b=0.056$ estimated with Geant4.
The neutron efficiency of the calorimeter
as a function of the energy and the scattering angle is also studied
with Geant4 in details. The calorimeter's responses to different
types of particles are carefully studied,
in order to have a realistic fast simulation of the spectrometer.

To understand the physics impact of the measurement, the statistic of the produced $\eta$
samples is the most important input for the simulation.
To be conservative on our experimental projections,
in this simulation, we consider a prior experiment of only one-month running.
Based on the evaluated luminosity and $p$-$A$ cross section,
the potential production rate of $\eta$ can be above $10^8$ s$^{-1}$,
with the inelastic event rate around $10^{10}$ s$^{-1}$.
The silicon pixel detector of high granularity is capable of working
at a high event rate ($>100$ MHz) without significant pile-ups of events.
Nevertheless, considering the radiation hardness of the detector
and the limit of current data acquisition (DAQ) system,
we make a very conservative estimation of the event rate for
Huizhou $\eta$ factory experiment. The event rate of inelastic scattering is
assumed to 100 MHz, and the $\eta$ production rate is about 760 KHz.
We also assume a conservative duty factor of the accelerator,
which is 30\%. With these settings, the number of $\eta$ mesons produced is
$5.9\times10^{11}$ for the very first experiment of only one-month running time.
Thus in the following simulations, we assume only $5.9\times10^{11}$
eta mesons produced in the prior experiment.

The statistic of $\eta$ meson samples can be increased magnitudes higher,
since the experiment will run for years, the event rate can also be increased
with the improvements of detector radiation hardness and speed of DAQ system,
and the proton beam can be delivered to the high-energy terminal with a high duty factor.

\subsection{Dark photon search}

\begin{figure}[htbp]
\begin{center}
  \includegraphics[width=0.47\textwidth]{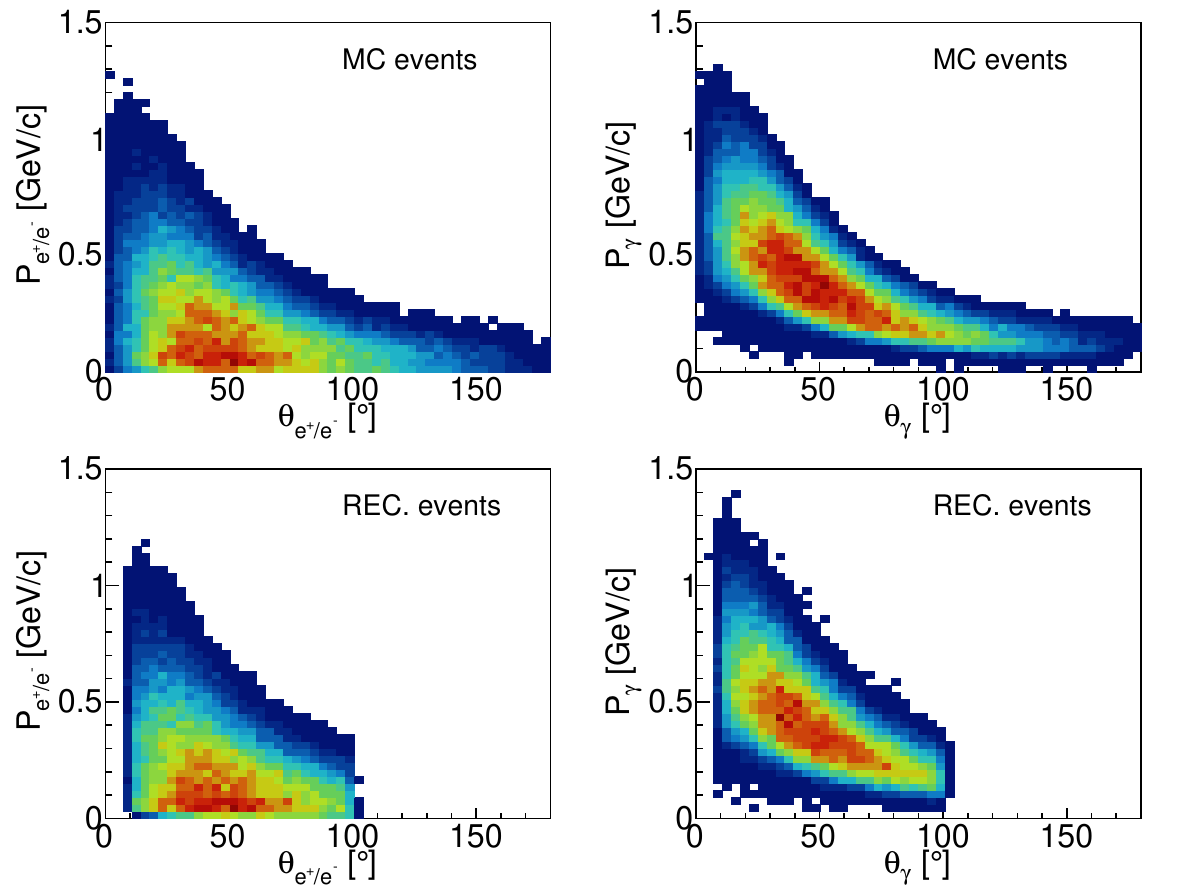}
  \caption{
    The momentum v.s. angle distributions of the final-state particles of $\eta\rightarrow e^+e^-\gamma$ decay channel.
    The top pads show the kinematic distributions of the final states from the event generator,
    while the bottom pads show the the kinematic distributions of the reconstructed particles from the fast detector simulation.
    The designed spectrometer covers the main and a large kinematic region of the final-state particles.
  }
  \label{fig:eta_decay_kinematics_EEG-channel}
\end{center}
\end{figure}

\begin{figure}[htbp]
\begin{center}
  \includegraphics[width=0.47\textwidth]{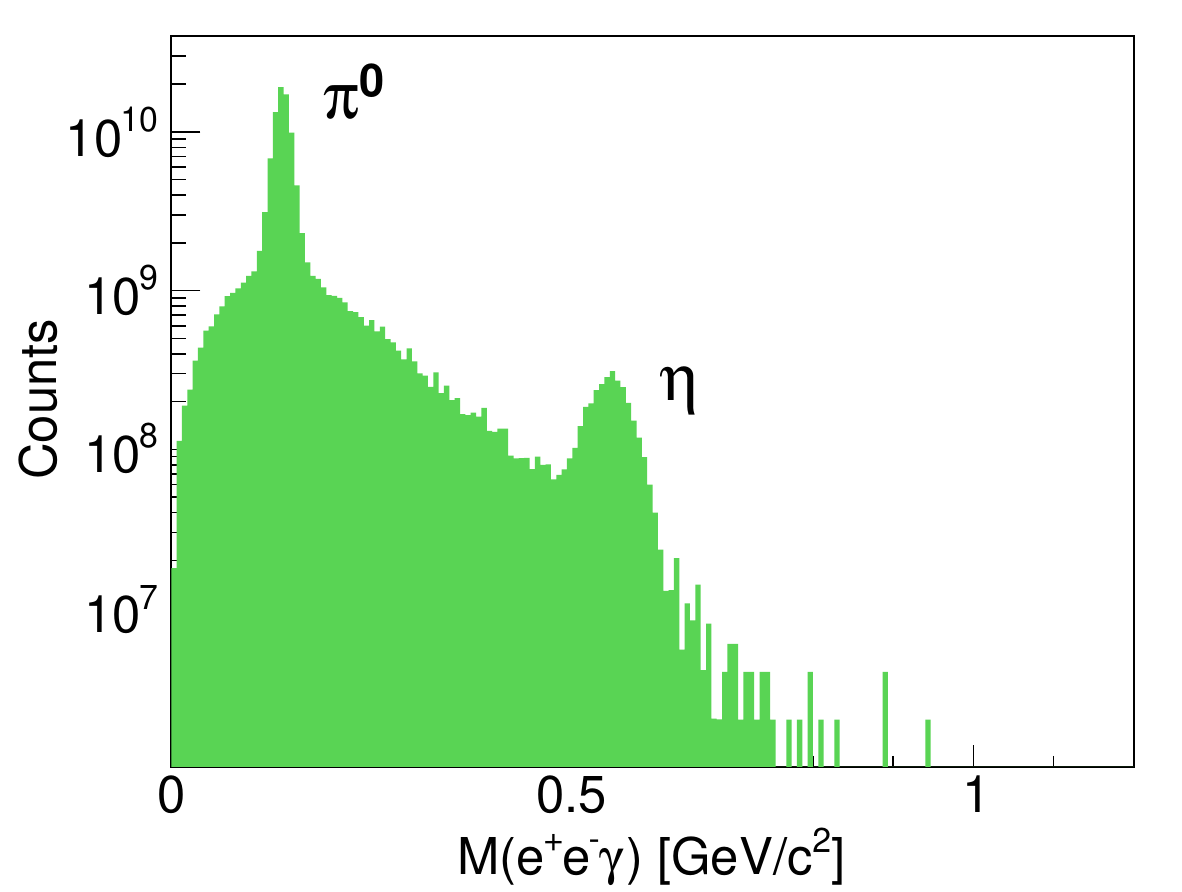}
  \caption{
    The invariant mass distribution of $e^+e^-\gamma$ from the simulation data
    for one-month running of Huizhou $\eta$ factory experiment.
  }
  \label{fig:eta_mass_distribution_eegchannel}
\end{center}
\end{figure}

\begin{figure}[htbp]
\begin{center}
  \includegraphics[width=0.47\textwidth]{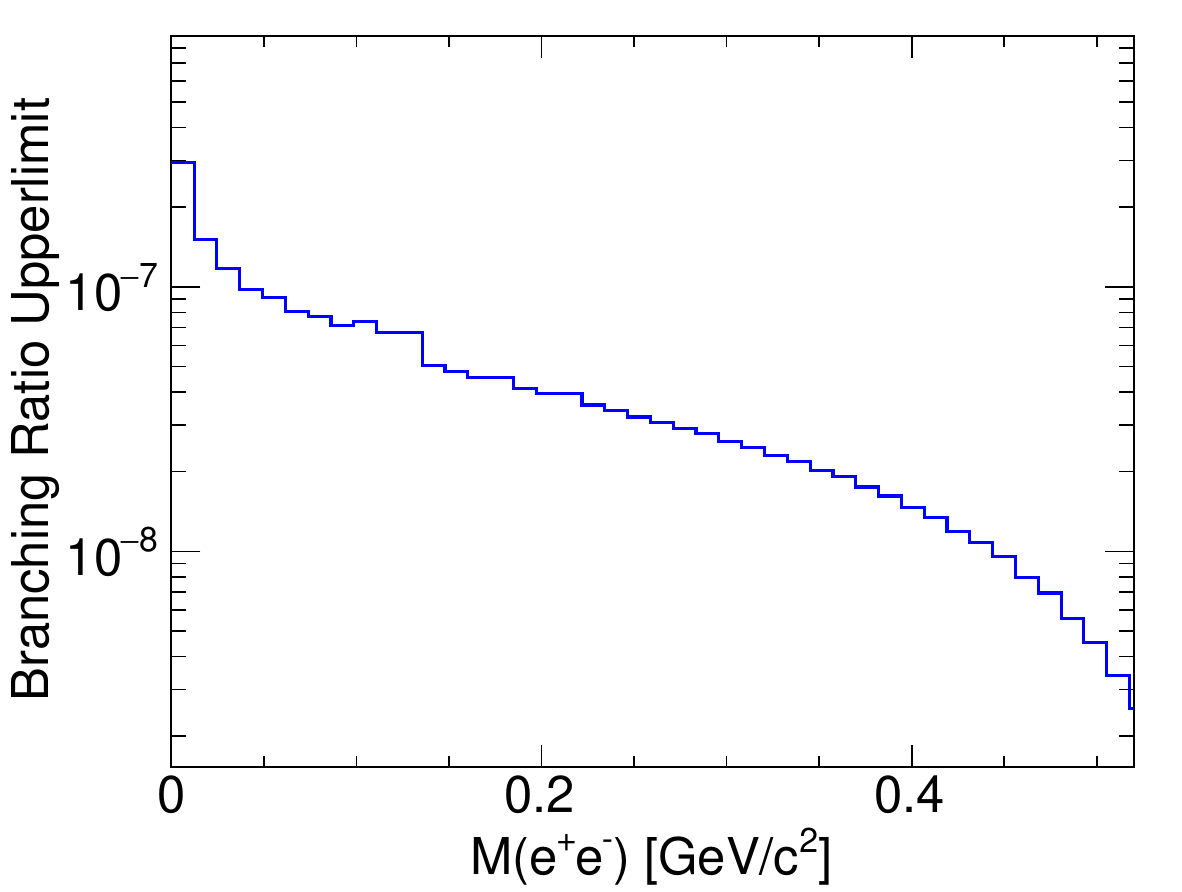}
  \caption{
    The estimated branching-ratio upperlimit of dark photon
    for just one-month running of Huizhou $\eta$ factory experiment,
    under a conservative event rate of 100 MHz of inelastic reactions.
  }
  \label{fig:dark_photon_BrRatio_uplimit}
\end{center}
\end{figure}

\begin{figure}[htbp]
\begin{center}
  \includegraphics[width=0.47\textwidth]{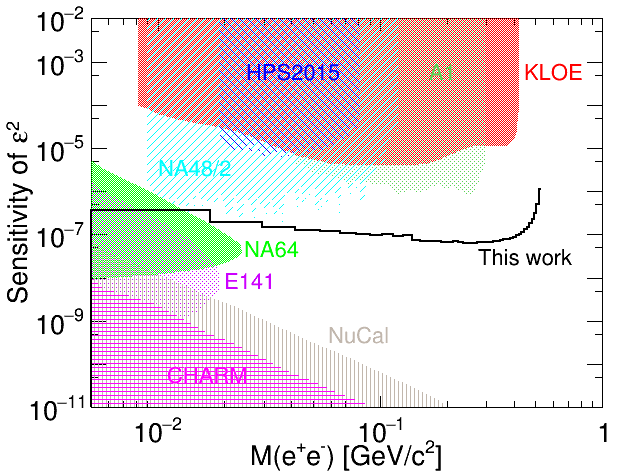}
  \caption{
    The estimated $\varepsilon^2$ sensitivity of dark photon,
    for just one-month running of Huizhou $\eta$ factory experiment,
    under a conservative event rate of 100 MHz of inelastic reactions.
    The shaded exclusion areas in the figure
    of previous experiments (HPS2015, A1@MAMI, KLOE, NA48/2, NA64, E141, NuCal and CHARM)
    are taken from Refs. \cite{HPS:2018xkw,A1:2011yso,Giovannella:2011nh,NA482:2015wmo,Gninenko:2300189,
    Riordan:1987aw,Blumlein:2011mv,Blumlein:2013cua,Gninenko:2012eq}.
  }
  \label{fig:dark_photon_epsilon2_sensitivity}
\end{center}
\end{figure}

The decay channel $\eta\rightarrow e^+ e^- \gamma$ is of particular interests,
as this channel is relevant for the search of the dark photon
\cite{Holdom:1985ag,Galison:1983pa,Fayet:1990wx,Fayet:1980rr}
and the light protophobic X17 boson \cite{Krasznahorkay:2015iga,Feng:2016jff,Feng:2016ysn}
which decay into the electron-positron pair.
At the same time, from the precise measurement of this channel,
we can precisely extract the transition form factor of $\eta$,
which is an important input for the theoretical calculation of muon anomalous moment $(g-2)_{\mu}$.
The dark photon is the most popular dark portal particle
which feebly connects the SM model sector with the possible hidden sector.
Here we focus on the physics impact on the dark photon
from the simulation data of Huizhou eta factory experiment.

Fig. \ref{fig:eta_decay_kinematics_EEG-channel} shows the kinematic distributions of
the final-state particles of the channel $e^+ e^- \gamma$,
from the event generator and the particle reconstruction of the spectrometer simulation.
One sees that the majority of the final electrons are of low momentum ($<0.5$ GeV/c)
and go to a angle from 10 $^{\circ}$ to 100 $^{\circ}$.
The average energy of the final photon is about 0.4 GeV,
and the photons are of the similar scattering angles of the electrons.
The designed spectrometer just covers the most of the electrons and photons,
the overall efficiency for the channel is estimated to be 60\% with the simulation.
For the low-energy electron, it can be effectively identified with
the energy decomposition $dE/dx$ measured by the silicon pixel tracker.
And for the high-energy electron, it can be identified with the
help of the calorimeter, as the pion initiates very few Cherenkov photons
in the lead glass calorimeter.

The distribution of reconstructed invariant mass of $e^+e^-\gamma$ is
shown in Fig. \ref{fig:eta_mass_distribution_eegchannel}.
One sees clearly the peaks of $\pi^0$ and $\eta$
with just very low background underneath.
Owing to the suppression of bremsstrahlung radiations
in the proton scattering process, the electron and photon
backgrounds are not significant.
The $\eta$ samples can be selected with a high purity by performing
a cut on the invariant mass of $e^+e^-\gamma$.
In this simulation study, the invariant mass is
required to be in the range of $[m_{\eta}-3\sigma,m_{\eta}+3\sigma]$.

To estimate the sensitivity of the proposed experiment to the dark photon,
we carefully studied the background distribution with the simulation.
The background events are generated with GiBUU and with some
decay chains added by us. In the simulation data, there is no bump
in the invariant mass distribution of electron and position.
We assume there is no dark photon in the simulation and
the invariant mass distribution of $e^+ e^-$ is the pure background distribution.
No observation of the dark photon means that the statistical
significance of the dark photon peak is smaller than $3\sigma$.
As a result, we get a formula for the branching-ratio (BR) upper limit of
the dark photon channel, which is given by,
\begin{equation}
  \begin{split}
    {\rm BR^{up}} = \frac{3\times \sqrt{N_{\rm bg}\times \epsilon_{\rm bg}}}
    {N_{\rm\eta}\times \epsilon_{\rm sig}},
  \end{split}
  \label{eq:BrUplimit}	
\end{equation}
where $N_{\rm bg}$ is the number of background events,
$\epsilon_{\rm bg}$ is efficiency for the background event,
$N_{\rm\eta}$ is the total number of eta mesons produced in the experiment,
and $\epsilon_{\rm sig}$ is the efficiency for the dark photon channel.
$N_{\rm bg}\times \epsilon_{\rm bg}$ is actually the number of background events
survived after all the event selections.
Based on the simulation of one-month experiment of Huizhou eta factory,
the BR upper limit of dark photon in $\eta$ decay is evaluated
and shown in Fig. \ref{fig:dark_photon_BrRatio_uplimit}.
The sensitivity of kinematic mixing parameter $\epsilon^2$ is closely
related to the upper limit of branching ratio, which is written as,
\begin{equation}
  \begin{split}
    S(\epsilon^2) = \frac{\rm BR^{up}}
    {2|F(m_{\rm A}^2)|^2\left(1-\frac{m_{\rm A}^2}{m_{\rm\eta}^2}\right)^3},
  \end{split}
  \label{eq:ParSensitivity}	
\end{equation}
in which $m_{\rm A}$ and $m_{\rm\eta}$ are the masses of the dark photon
and the $\eta$ meson respectively, and $F$ is the transition form factor of $\eta$.
The final $\epsilon^2$ sensitivity of one-month experiment to the dark photon is shown
in Fig. \ref{fig:dark_photon_epsilon2_sensitivity}.
Our simulation indicates a significant sensitivity to $\epsilon^2$, below $10^{-7}$,
which surpass the precisions of the previous experimental measurements
(HPS2015 \cite{HPS:2018xkw}, A1@MAMI \cite{A1:2011yso}, KLOE \cite{Giovannella:2011nh},
NA48/2 \cite{NA482:2015wmo}, NA64 \cite{Gninenko:2300189},
E141 \cite{Riordan:1987aw}, NuCal \cite{Blumlein:2011mv,Blumlein:2013cua}
and CHARM \cite{Gninenko:2012eq}).
The proposed experiment will be a valuable compliment to other dark photon searches.
With years running of the experiment, the parameter space below the $\eta$ mass
will be almost ruled out combined with many other experiments worldwide
\cite{Adrian:2022nkt,HPS:2018xkw,A1:2011yso,APEX:2011dww,Bjorken:2009mm,Marsicano:2018krp,Batell:2014mga,
Blumlein:2013cua,Gninenko:2012eq,Blumlein:2011mv,Bjorken:1988as}.

\subsection{Light dark higgs search}

\begin{figure}[htbp]
\begin{center}
  \includegraphics[width=0.47\textwidth]{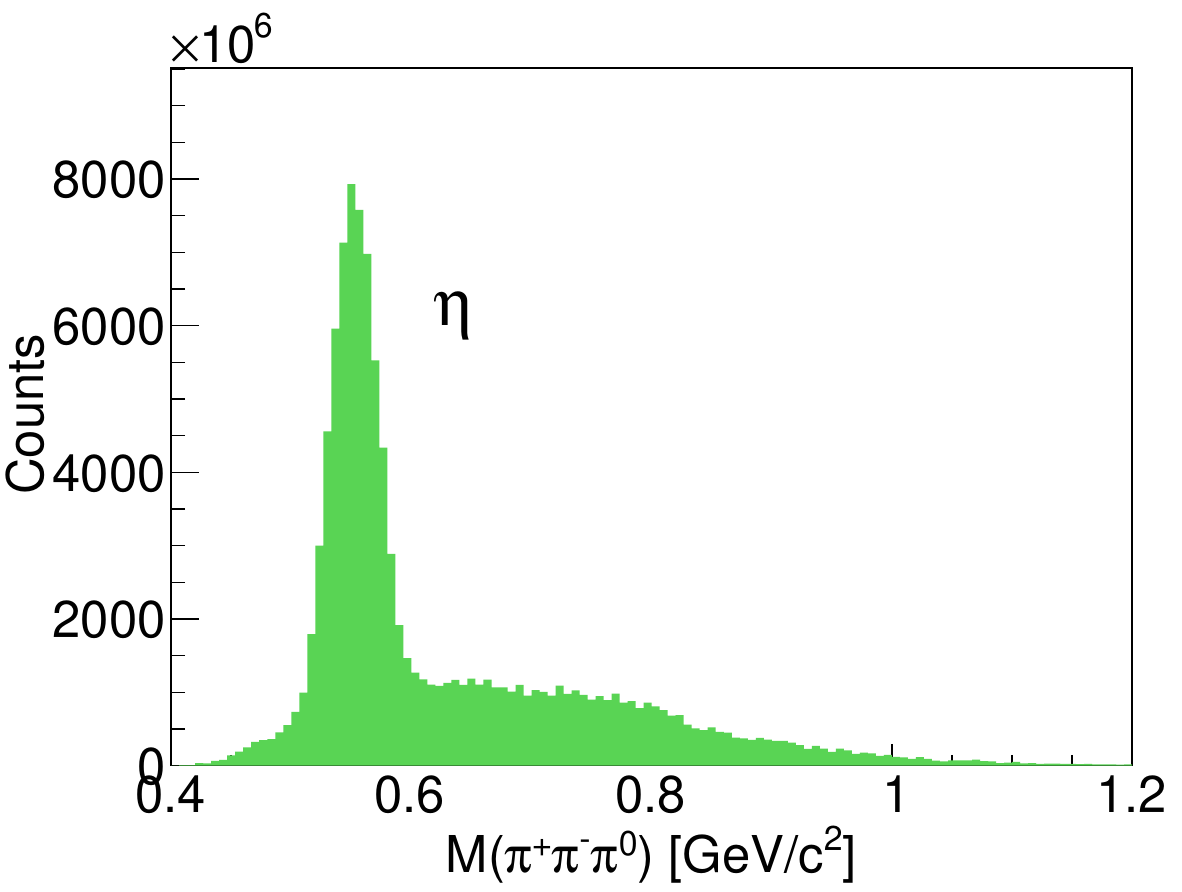}
  \caption{
    The invariant mass distribution of $\pi^+\pi^-\pi^0$ from the simulation data
    for one-month running of Huizhou $\eta$ factory experiment.
  }
  \label{fig:eta_mass_distribution_3pichannel}
\end{center}
\end{figure}

\begin{figure}[htbp]
\begin{center}
  \includegraphics[width=0.47\textwidth]{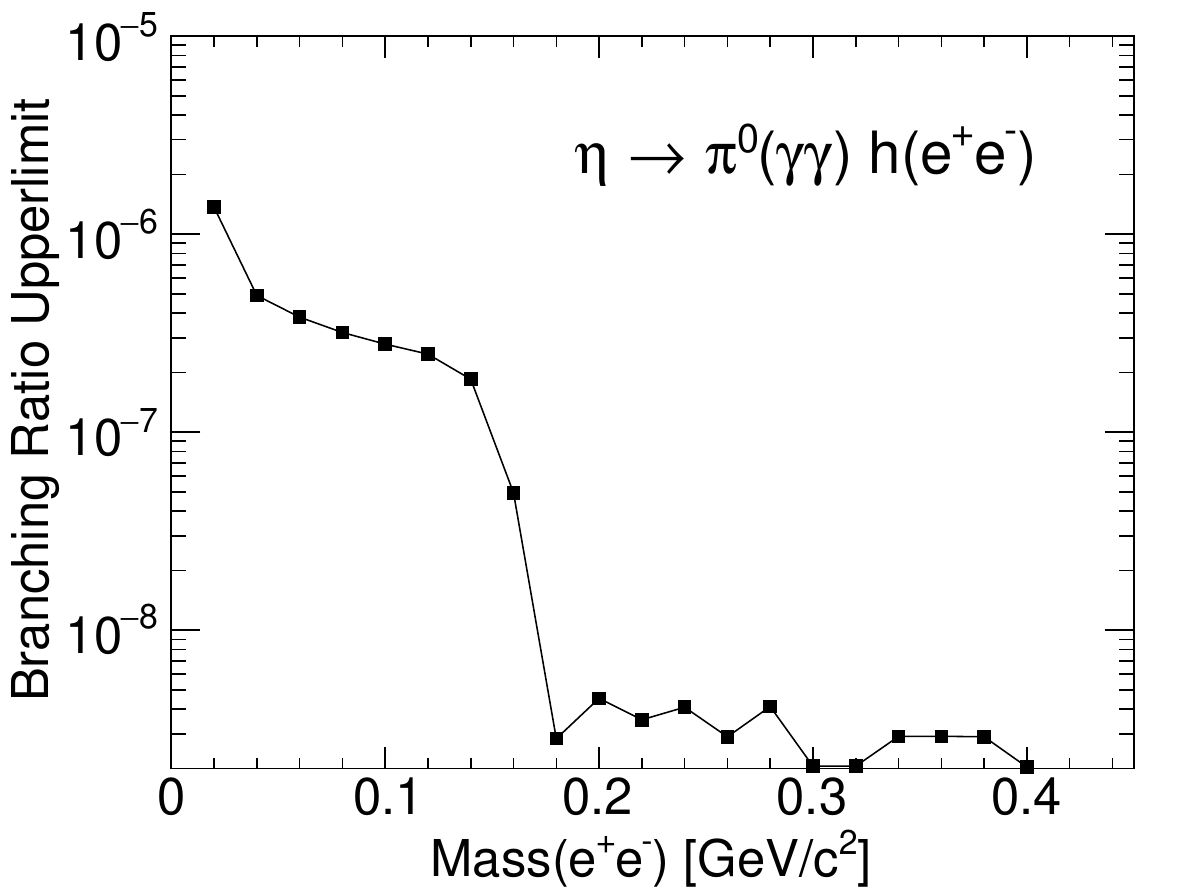}
  \caption{
    The estimated branching-ratio upperlimit of light dark higgs particle
    from $\pi^0 e^+ e^-$ channel
    for just one-month running of Huizhou $\eta$ factory experiment,
    under a conservative event rate of 100 MHz of inelastic reactions.
    The invariant mass of $\pi^0 e^+ e^-$ is required to be in the $\eta$ mass region.
  }
  \label{fig:BR-uplimit-dark-higgs-electron-channel}
\end{center}
\end{figure}

\begin{figure}[htbp]
\begin{center}
  \includegraphics[width=0.47\textwidth]{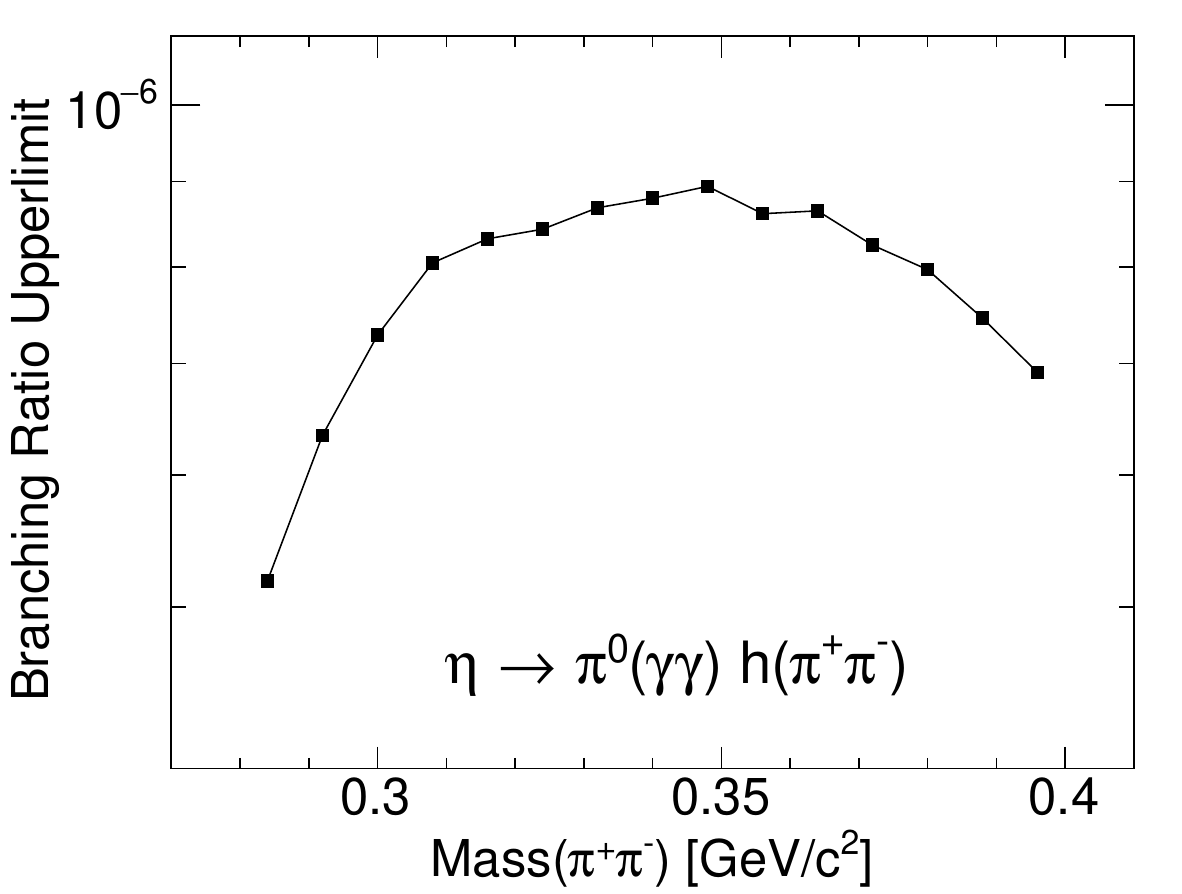}
  \caption{
    The estimated branching-ratio upperlimit of light dark higgs particle
    from $\pi^0 \pi^+ \pi^-$ channel
    for just one-month running of Huizhou $\eta$ factory experiment,
    under a conservative event rate of 100 MHz of inelastic reactions.
    The invariant mass of $\pi^0 \pi^+ \pi^-$ is required to be in the $\eta$ mass region.
  }
  \label{fig:BR-uplimit-dark-higgs-pion-channel}
\end{center}
\end{figure}

\begin{figure}[htbp]
\begin{center}
  \includegraphics[width=0.47\textwidth]{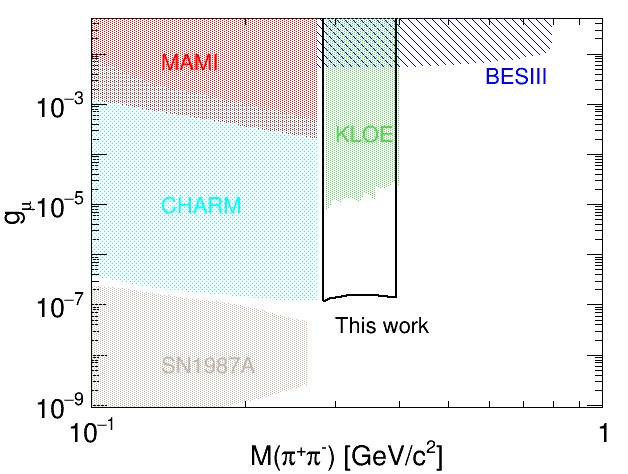}
  \caption{
    The estimated $g_u$ sensitivity of light dark higgs particle
    in a hadrophilic scalar model \cite{Batell:2017kty,Batell:2018fqo},
    for just one-month running of Huizhou $\eta$ factory experiment,
    under a conservative event rate of 100 MHz of inelastic reactions.
    The previous experimental data for the shaded exclusion areas
    (BESIII, KLOE, MAMI, CHARM, and SN1987A) in the figure
    are taken from Refs. \cite{BESIII:2016tdb,KLOE-2:2016zfv,
    A2atMAMI:2014zdf,CHARM:1985anb,Liu:2018qgl,Batell:2018fqo}.
  }
  \label{fig:dark_scalar_gu_sensitivity}
\end{center}
\end{figure}

The light dark higgs
\cite{Burgess:2000yq,OConnell:2006rsp,Batell:2018fqo,Batell:2017kty,Patt:2006fw,Silveira:1985rk,Pospelov:2007mp}
is another representative dark portal particle,
which couples the hidden scalar field with the Higgs doublet.
The dark higgs is thus weakly connected with the leptons and quarks
via the Yukawa coupling. Therefore the dark higgs can be produced in the hadronic
process and it can decays into lepton pairs and quark pairs.
In a hadrophilic scalar model, the dark higgs mainly couples to the up quark,
thus it dominantly decays into pions.
At Huizhou eta factory, we could search for the dark higgs in the following channels:
$\eta\rightarrow \pi^0 h\rightarrow \pi^0 e^+ e^-$ and
$\eta\rightarrow \pi^0 h \rightarrow \pi^0\pi^+\pi^-$.
In these $\eta$ rare decay channels, a bump in the invariant mass distribution
of $e^+ e^-$ or $\pi^+ \pi^-$ is a clear signal of the possible dark scalar particle.

The distribution of reconstructed invariant mass of $\pi^+\pi^-\pi^0$ is
shown in Fig. \ref{fig:eta_mass_distribution_3pichannel}.
One sees clearly the peak of $\eta$ meson
with a low background underneath.
In GiBUU simulation, the background from the direct multi-pion
production is low compared to the $\eta$ production,
as the incident energy of proton is not large (1.8 GeV).
The $\eta$ samples from $\pi^+\pi^-\pi^0$ can be selected
with a high purity by performing
a cut on the invariant mass of $\pi^+\pi^-\pi^0$ in the range
of $[m_{\eta}-3\sigma,m_{\eta}+3\sigma]$.
The low background does not hinder much our explorations
for the rare decays of $\eta$ meson.

From the simulation, the efficiencies of $\pi^0 e^+ e^-$ channel and
$\pi^0\pi^+\pi^-$ channel are all above 40\% with the conceptual design of spectrometer.
The resolutions for invariant masses of $e^+ e^-$ and $\pi^+\pi^-$ are
2 MeV/c$^2$ and 1 MeV/c$^2$ respectively.
In our studies, the bin width for the invariant mass is six times of the resolution.
The background distributions without the dark higgs particle are simulated
with the GiBUU event generator, and the total number of inelastic scattering events scales
up to $5.9\times10^{11}$.
Since there is no dark higgs observed in our simulation data,
the upper limit of branching ratio of the dark higgs particle is given
with the formula in Eq. (\ref{eq:BrUplimit}).
The BR upper limits of the light dark higgs particle in
$\pi^0 e^+ e^-$ and $\pi^0\pi^+\pi^-$ channels are shown
in Fig. \ref{fig:BR-uplimit-dark-higgs-electron-channel}
and Fig. \ref{fig:BR-uplimit-dark-higgs-pion-channel} respectively,
as a function of the mass of the dark higgs.

From Fig. \ref{fig:BR-uplimit-dark-higgs-pion-channel},
one sees that BR upper limit of dark higgs
in $\eta\rightarrow\pi^0\pi^+\pi^-$ channel is at the level of $10^{-6}$ to $10^{-7}$,
for one-month running of the experiment.
From Fig. \ref{fig:BR-uplimit-dark-higgs-electron-channel},
one sees that the upper limit in $\pi^0 e^+ e^-$ channel is obviously
below $10^{-8}$ in the most range of dark higgs mass.
This is mainly due to the less electron background in $p-A$ collisions,
comparing to the strong pion background.
One also finds that the upper limit in $e^+ e^-$ channel falls down quickly,
reaching a value below $10^{-8}$ when the mass is above 0.14 GeV.
This is because most of the $e^+ e^-$ background comes from the $\pi^0$ decay.
Therefore the $e^+ e^-$ channel has the advantage in searching
the dark higgs of the mass heavier than the pion mass.
With years running of the Huizhou eta factory experiment,
we are confident that the accumulated data would provide the strong
constraints of the possible dark higgs particle.

Under the hadrophilic scalar model \cite{Batell:2017kty,Batell:2018fqo},
the sensitivity to the parameter $g_u$
(the coupling of dark scalar to the first-generation quark) is computed
and shown in Fig. \ref{fig:dark_scalar_gu_sensitivity},
compared with the constraints provided by the previous experimental data
(BESIII \cite{BESIII:2016tdb}, KLOE \cite{KLOE-2:2016zfv},
MAMI \cite{A2atMAMI:2014zdf}, CHARM \cite{CHARM:1985anb,Liu:2018qgl} and
SN1987A \cite{Batell:2018fqo}).
One finds that the $g_u$ sensitivity from one-month running of the proposed
Huizhou $\eta$ factory will exceed the current experimental limits
in the accessed mass range. The proposed super $\eta$ factory will play
an important role in searching the light dark scalar portal particles.

\subsection{C and CP violation in $\eta\rightarrow \pi^+\pi^-\pi^0$}

\begin{figure}[htbp]
\begin{center}
  \includegraphics[width=0.47\textwidth]{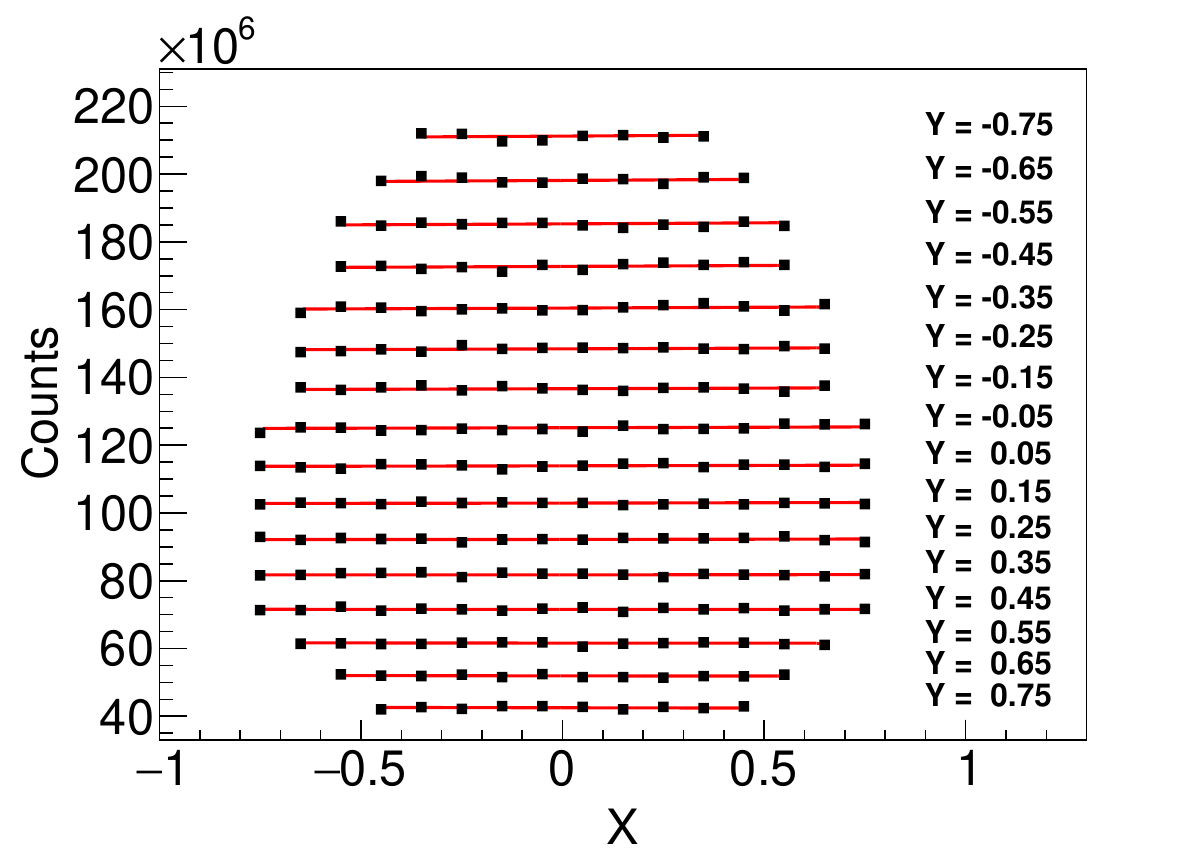}
  \caption{
    The event distributions of $\eta\rightarrow \pi^+\pi^-\pi^0$ decay channel (black squares)
    in different $X$ and $Y$ bins,
    for just one-month running of Huizhou $\eta$ factory experiment,
    under a conservative event rate of 100 MHz of inelastic reactions.
    The Dalitz distribution of $\eta\rightarrow \pi^+\pi^-\pi^0$ is fitted
    with a simple model (the red lines).
    See the main text for more explanations.
  }
  \label{fig:fitting_eta_decay_Dalitz_3pi}
\end{center}
\end{figure}

The CP violation in the flavor-nondiagonal process due to
the Cabibbo-Kobayashi-Maskawa (CKM) matrix phase is not enough to
explain the matter-antimatter asymmetry in the universe.
Therefore, searching for new sources and flavor-diagonal
CP violation has been a hot topic in high-energy physics.
The $\pi^+\pi^-\pi^0$ decay channel of $\eta$ meson is of particular interests,
since it provides a unique process to probe the flavor-diagonal
C and CP violation beyond the SM.
This type of CP violation is not constrained by the measurement
of nucleon Electro-Dipole Moment (EDM), thus it lacks of
the experimental study of high precision \cite{Gardner:2019nid}.
Owing to the interference between the C-conserving and C-violating amplitudes,
the CP violation signal can be large.
It is possible to find the small C and CP violation from a precise measurement
of the mirror symmetry in the Dalitz decay plot of $\pi^+\pi^-\pi^0$ channel.

The direct observable of the charge asymmetry and CP violation is the mirror symmetry
breaking in the Dalitz plot of $\eta\rightarrow \pi^+\pi^-\pi^0$,
i.e. the asymmetry under the exchange of $u$ and $t$
($u\equiv (p_{\pi^+}+p_{\pi^0})^2$, $t\equiv (p_{\pi^-}+p_{\pi^0})^2$,
and $s\equiv (p_{\pi^+}+p_{\pi^-})^2$).
The C and CP violation is reflected in the asymmetry of
the decay events of $u>t$ and $u<t$.
Usually, the mirror asymmetry is vividly illustrated in the Dalitz plot
of X and Y variables, which are defined as,
\begin{equation}
  \begin{split}
    X\equiv \sqrt{3}\frac{T_{\pi^+}-T_{\pi^-}}{Q_{\eta}}
    =\frac{\sqrt{3}}{2m_{\eta}Q_{\eta}}(u-t), \\
    Y\equiv \frac{3T_{\pi^0}}{Q_{\eta}} - 1
    =\frac{3}{2m_{\eta}Q_{\eta}}[(m_{\eta}-m_{\pi^0})^2 - s] - 1,
  \end{split}
  \label{eq:X-Y-defs}	
\end{equation}
where $Q_{\eta}=m_{\eta}-m_{\pi^+}-m_{\pi^-}-m_{\pi^0}$ and $T_{\pi^i}$ are the total
kinematic energy and the kinematic energy of $\pi^i$ in the $\eta$ rest frame.
The distribution asymmetry across $X=0$ is a observable of the new type of CP violation.
The Dalitz distribution of the decay probability can be conveniently
parameterized as a polynomial expansion, which is written as,
\begin{equation}
  \begin{split}
    N(X,Y) = N_0(1+aY+bY^2+cX+dX^2+eXY \\
    +fY^3+gX^2Y+hXY^2+lX^3+...),
  \end{split}
  \label{eq:dalitz-parametrization}	
\end{equation}
in which $a,b,c...$ are free parameters.
The nonzero value of the parameter $c$, $e$, $h$, or $l$ is a strong indication
of the flavor-diagonal C and CP violation.

The $\pi^+\pi^-\pi^0$ channel is one of the major decay channels of $\eta$ meson,
and we can get a huge amount of the decay events from Huizhou eta factory experiment.
From the simulation, the efficiency for the 3 pion channel is estimated
to be approximate 45\%. The event distributions in different X and Y bins are shown
in Fig. \ref{fig:fitting_eta_decay_Dalitz_3pi}, for one-month running of the experiment.
The statistical error bars are too small to appear in the figure.
We performed a model fit to the data with Eq. (\ref{eq:dalitz-parametrization}).
The uncertainty of the parameter $c$ is about $5\times 10^{-5}$,
which is two magnitudes smaller than the current analyses of COSY and
KLOE-II data \cite{KLOE-2:2016zfv,WASA-at-COSY:2014wpf}.
With years running of the project, the C and CP violation
can be tested at a splendid level of precision.

\subsection{Low-background $\eta$ data from exclusive channel $pd\rightarrow\eta~{\rm ^3He}$}

\begin{figure}[htbp]
\begin{center}
  \includegraphics[width=0.47\textwidth]{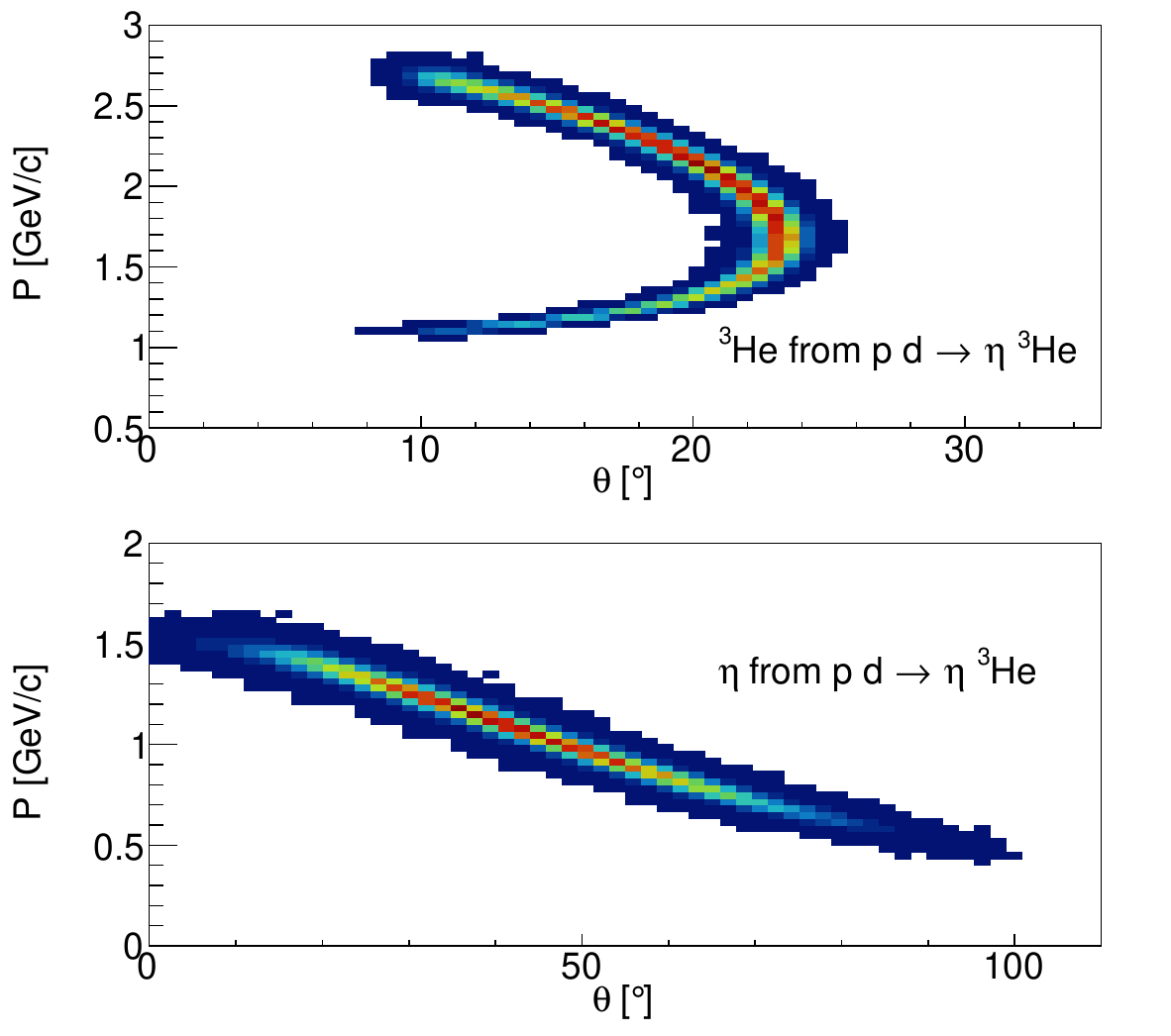}
  \caption{
    The kinematic distributions of the reconstructed $^3$He and $\eta$ from
    a fast simulation of the spectrometer.
    The scattering angle and the momentum are highly correlated for
    the particles in the reaction $pd\rightarrow\eta~{\rm ^3He}$.
    The angular and momentum resolutions are small from the silicon pixel tracker.
  }
  \label{fig:eta_3He_kinematics_pd-channel}
\end{center}
\end{figure}

We emphasis here that the low-background data of $\eta$ mesons can be achieved
at Huizhou eta factory via the $^3$He tagged events of the reaction $pd\rightarrow\eta~{\rm ^3He}$.
In addition to the exclusivity of the measurement,
the momentum and angle of the final-state particle are highly correlated
in the two-body-to-two-body scattering process.
By tagging $^3$He and a cut of the momentum-angle correlation,
the background is enormously reduced.
The cross section of $pd\rightarrow\eta~{\rm ^3He}$ reaction is not small
\cite{Wilkin:2016mfn,Bilger:2002aw,Smyrski:2007nu,Mersmann:2007gw,Rausmann:2009dn,WASA-at-COSY:2014xva},
which is 0.4 $\mu$b near the production threshold measured
by COSY-ANKE collaboration \cite{Mersmann:2007gw}.
Note that the multiplicity of the final particles using deuterium target is
much smaller than that using other nuclear targets.
Therefore the event rate for $pd\rightarrow\eta~{\rm ^3He}$ measurement
can be set at a much higher rate, to increase the statistic
of the low-background data.

Fig. \ref{fig:eta_3He_kinematics_pd-channel} shows the two-dimensional kinematic
distributions of $^3$He and $\eta$ in the momentum vs. angle plane.
One sees that the final $^3$He mainly goes to the region of
scattering angle from 15$^\circ$ to 25$^\circ$,
while the $\eta$ meson has the scattering angle mainly in the range
from 20$^\circ$ to 70$^\circ$.
The conceptual design of the spectrometer is just suitable
for tagging $^3$He and collecting the decay particles of $\eta$ meson
with a large acceptance. From Fig. \ref{fig:eta_3He_kinematics_pd-channel},
one sees that the momentum and angular resolutions of the silicon pixel tracker are
excellent in picking up the exclusive events of $pd\rightarrow\eta~{\rm ^3He}$.

In short, with the high-intensity proton beam and the deuterium target,
we could have a measurement of both high luminosity and high precision
at Huizhou eta factory. This high-statistic and low-background data is preciously
valuable in searching the new light particles,
looking for the violations of CP and other discrete symmetries,
measuring the transition form factor and $u-d$ quark mass difference,
and testing the low-energy effective theory of strong interaction.
The systematic uncertainty from the background can be well controlled with
the tagged $\eta$ data of $pd\rightarrow\eta~{\rm ^3He}$.

\section{Summary and outlook}
\label{sec:summary}

A super $\eta$ factory at Huizhou is proposed to pursue a variety of
meaningful and challenging physical goals.
The HIAF accelerator complex and the conceptual design of the spectrometer
are briefly discussed. More than $10^{13}$ $\eta$ mesons can be produced
with 100\% duty factor of the accelerator.
The performance of the spectrometer is studied with Geant4 simulation,
demonstrating satisfactory efficiency and resolution.
The designed spectrometer is particularly adept at detecting charged particles
and exhibits the required radiation hardness for high-luminosity experiments.

Through simulations, some key channels of Huizhou $\eta$ factory experiment are investigated.
The preliminary results from the fast simulation manifest that the Huizhou $\eta$ factory will
play a crucial role in searching for the predicted light dark portal particles
and new sources of CP violation.
The proposed experiment has the potential to significantly constrain the parameter space
of the dark photon in the low-mass region together with other experiments.
The sensitivity to the light dark scalar particle is estimated to be at an unprecedented level.
The C and CP violation in channel $\eta\rightarrow \pi^+\pi^-\pi^0$ can be measured
by at least two magnitudes more precise, comparing to the up-to-date measurement worldwide.
Judged with the simulation, the conceptual design of the spectrometer is capable for
measuring the tagged $\eta$ events of the reaction $pd\rightarrow\eta~{\rm ^3He}$.
The tagging $^3$He method provides a measurement of both high statistic and low background,
which is vital for the precision study of $\eta$ physics.

Upon completing the planned accumulation of $\eta$ decay samples,
we could increase the beam energy and produce the $\eta^{\prime}$ mesons.
The physical goals of high-precision studies of $\eta^{\prime}$ meson decays closely resemble
to those of $\eta$ meson. An advantage of studying $\eta^{\prime}$ decay
is the ability to explore dark portal particles in a wider mass range,
given that $\eta^{\prime}$ meson is heavier than $\eta$ meson.
With the same spectrometer, we can also carry out high-precision studies
on $\eta^{\prime}$ and $\phi$ meson decays, thereby boosting the discovery potential
of the proposed Huizhou $\eta$ factory project.

To further improve the discovery potential of the spectrometer,
it is essential to enhance its capacity for detecting neutral particles.
The current lead-glass EM calorimeter exhibits standard energy resolution;
therefore exploring new calorimeter technologies with fast response time ($<100$ ps)
and small energy resolution ($<3.5$\% at 1 GeV) is imperative.
With the rapid developments of silicon photomultiplier and electronics,
dual-readout calorimetry collecting scintillation photons and Cherenkov photons
could be a viable option for updating the EM calorimeter.
The scintillation material improves the energy resolution significantly,
while the Cherenkov light gives the sharp time resolution.
With the dual readout calorimeter, the ability of particle identification also can be enhanced
by comparing scintillation signal amplitude and Cherenkov signal amplitude.
Future developments in silicon pixel detector and electronics will benefit
the proposed Huizhou $\eta$ factory project, enabling improvements in radiation hardness
and resolutions that will elevate the event-rate limit for the planned high-luminosity experiments.

\section*{Acknowledgments}
This work was supported by the National Natural Science Foundation
of China under the Grant NOs.  12222512 and 12005266,
the Strategic Priority Research Program of Chinese Academy of
Sciences under the Grant NO. XDB34030300.

\bibliographystyle{unsrt}
\bibliography{references}

\end{document}